\documentclass[aps,floatfix,showpacs,nofootinbib,showkeys,twocolumn,superscriptaddress]{revtex4-1}
\usepackage[T1]{fontenc}
\usepackage{amssymb,amsfonts,amsmath,mathtext,enumerate,float,mathrsfs}
\usepackage[utf8]{inputenc}%включаем свою кодировку: koi8-r или utf8 в UNIX, cp1251 в Windows
\usepackage[english]{babel}%используем русский и английский языки с переносами
\usepackage{graphicx,color}
\usepackage{wrapfig}
\graphicspath{{./}}

\usepackage{epsfig}
\usepackage{epsf}
\usepackage{subfig} % for figures in a number
\usepackage{feynmp} % for to create diagram Feyman
\usepackage{pgf,pgfarrows,pgfnodes,pgfautomata,pgfheaps,multirow}
\usepackage{array}
\usepackage{bm}% bold math \
\usepackage{latexsym}
\usepackage{pifont}

\newcommand{\be}{\begin{equation}}
\newcommand{\ee}{\end{equation}}
\newcommand{\bea}{\begin{eqnarray}}
\newcommand{\eea}{\end{eqnarray}}
\newcommand\nn{\nonumber}

\begin{document}

%\journalname{prc}

\title{The decay $\tau \to K^{*-}(892) \eta \nu_{\tau}$ in the NJL model}

\author{M.\ K.\ Volkov}
\email{volkov@theor.jinr.ru}
\affiliation{Bogoliubov Laboratory of Theoretical Physics, JINR, Dubna, 141980 Russia}

\author{A.\ A.\ Pivovarov}
\email{tex$\_$k@mail.ru}
\affiliation{Bogoliubov Laboratory of Theoretical Physics, JINR, Dubna, 141980 Russia}

\author{K.\ Nurlan}
\email{nurlan.qanat@mail.ru}
\affiliation{Bogoliubov Laboratory of Theoretical Physics, JINR, Dubna, 141980 Russia}
\affiliation{Institute of Nuclear Physics, Almaty, 050032, Republic of Kazakstan}
\affiliation{Eurasian National University, Nur-Sultan, 01008, Republic of Kazakstan}

\begin{abstract}

The decay width $\tau \to K^{*-}(892) \eta \nu_{\tau}$ was calculated in the Nambu--Jona-Lasinio model. Four channels were taken into account: the contact channel (the final states are directly producted from lepton current without any intermediate meson states), the axial-vector channel with two intermediate physical states $K_{1}(1270)$ and $K_{1}(1400)$, the vector channel with intermediate $K^{*}(892)$ meson and the pseudoscalar channel with intermediate $K$ meson. It is shown that the first two channels give the dominant contribution to the decay width. Of the remaining two channels, the pseudoscalar channel plays a more prominent role. The final result is in satisfactory agreement with experimental data. Prediction for the differential decay width is presented.
\end{abstract}

\date{\today}



\maketitle

\section{Introduction}

Hadronic decays of $\tau$ lepton are very important for studying QCD phenomena at low energies. Various decay modes including an $\eta$ meson represent a wide class of decays that are still poorly studied. One of the important tasks in this field is to test effective chiral theories. The $\tau$ decay amplitudes in these theories can be obtained on the basis of the $U(3) \times U(3)$ chiral-symmetric models. At the same time, the partial and differential widths of the processes can be described within the standard Nambu-Jona-Lasinio (NJL) model \cite{Nambu:1961tp, Eguchi:1976iz,Ebert:1982pk,Volkov:1984kq,Volkov:1986zb,Ebert:1985kz,Vogl:1991qt,Klevansky:1992qe,Volkov:1993jw,Hatsuda:1994pi,Ebert:1994mf,Volkov:2005kw}.

The process $\tau \to K^{*-}(892) \eta \nu_{\tau}$ has been repeatedly studied in various theoretical \cite{Li:1996md,Dai:2018thd} and experimental works \cite{Tanabashi:2018oca}. In the paper \cite{Li:1996md}, to describe the decay width of $\tau \to K^{*-}(892) \eta \nu_{\tau}$, the $U(3) \times U(3)$ chiral-symmetric model and the vector dominance model (VDM) were used. As a result, in \cite{Li:1996md} with the mixing angle for $\eta$ and $\eta^{'}$ mesons $\theta = -20^{\circ}$, the value for the partial width $Br(\tau \to K^{*-}(892) \eta \nu_{\tau}) = 1.01 \times 10^{-4}$ was obtained at experimental value $Br(\tau \to K^{*-}(892) \eta \nu_{\tau})_{exp} = (1.38 \pm 0.15) \times 10^{-4}$ \cite{Tanabashi:2018oca}.
 
This process was also studied in \cite{Dai:2018thd}. However, it was used there to fit model parameters from experimental data and calculate other decay modes.

In the present paper, in the framework of the standard $U(3) \times U(3)$ chiral-symmetric NJL model, the decay width of $\tau \to K^{*-}(892) \eta \nu_{\tau}$ is calculated and the differential distribution over the invariant mass of the meson pair $K^{*}(892) \eta$ is predicted. In this model the decay width $Br(\tau \to K^{*-}(892) \eta \nu_{\tau}) = 1.22 \times 10^{-4}$ is obtained for the mixing angle $\theta = -19^{\circ}$ \cite{Volkov:1998ax} of $\eta$ and $\eta^{'}$ mesons. 

An important feature of the present paper is taking into account the splitting of the axial-vector meson $K_{1A}$ into two physical states, due to the mixing of strange axial-vector mesons $K_{1A}$ from nonet $^{3}P_{1}$ and $K_{1B}$ from nonet $^{1}P_{1}$:
	\begin{eqnarray}
	\label{K1AK1B}
		K_{1A} & = & K_{1}(1270)\sin{\alpha} + K_{1}(1400)\cos{\alpha}, \nonumber\\
		K_{1B} & = & K_{1}(1270)\cos{\alpha} - K_{1}(1400)\sin{\alpha},
	\end{eqnarray}
where $\alpha=57^{\circ}$	\cite{Volkov:2019fyk, Volkov:2019yhy}. 

This effect was also considered in the papers \cite{Volkov:1986zb, Volkov:1984fr, Li:1996md, Suzuki:1993yc, Li:1995tv, Volkov:2019cja}. 

\section{Effective quark-meson Lagrangian}

In the NJL model, a fragment of the quark-meson Lagrangian of the interaction containing the mesons involved in the process under consideration has the form \cite{Volkov:1986zb}:
	
\begin{eqnarray}
&& \Delta L_{int} = 
\bar{q} \biggl[ \frac{g_{K_{1}}}{2} \gamma^{\mu}\gamma^{5} \sum_{j=\pm} \lambda_{j}^{K} K^{j}_{1\mu}  \nn \\ && \qquad
+\frac{g_{K^{*}}}{2} \gamma^{\mu} \sum_{j = \pm}\lambda_{j}^{K} K^{*j}_{\mu} + i g_{K} \gamma^{5} \sum_{j = \pm} \lambda_{j}^{K} K^{j}  \nn \\ && \qquad
+ i \sin(\bar{\theta}) g_{\eta_{u}} \gamma^{5}  \lambda_{u}^{\eta} \eta - i \cos(\bar{\theta}) g_{\eta_{s}} \gamma^{5}  \lambda_{s}^{\eta} \eta \biggl]q,
\end{eqnarray}	
	where $q$ and $\bar{q}$ are the fields of u, d, and s quarks with constituent masses $m_{u} \approx m_{d} = 280$~MeV, $m_{s} = 420$~MeV; $\lambda$ are linear combinations of the Gell-Mann matrices; $\bar{\theta} = \theta - \theta_{0}$, where $\theta = -19^{\circ}$ is the deviation from the ideal mixing angle $\theta_{0} = 35.3^{\circ}$ of mesons $\eta$ and $\eta^{'}$ \cite{Volkov:1998ax}.
	
The constants of the quark-meson interaction:
	
\begin{displaymath}
\label{Couplings}
g_{K_{1}} = g_{K^{*}} = \left(\frac{2}{3}I_{11}\right)^{-1/2},
\end{displaymath}
\begin{displaymath}
g_{K} =  \left(\frac{4}{Z_{K}}I_{11}\right)^{-1/2},
\end{displaymath}
\begin{displaymath}
g_{\eta_{u}} =  \left(\frac{4}{Z_{\eta_{u}}}I_{20}\right)^{-1/2},
\end{displaymath}
\begin{displaymath}
g_{\eta_{s}} =  \left(\frac{4}{Z_{\eta_{s}}}I_{02}\right)^{-1/2},
\end{displaymath}
	
where
	\begin{displaymath}
		 Z_{\eta_{u}} = \left(1 - 6\frac{m_{u}^{2}}{M^{2}_{f_{1}(1285)}}\right)^{-1}, 
	 \end{displaymath}
	 \begin{displaymath}
		 Z_{\eta_{s}} = \left(1 - 6\frac{m_{s}^{2}}{M^{2}_{f_{1}(1420)}}\right)^{-1},
	 \end{displaymath}
	 \begin{displaymath}
          Z_{K} = \left(1 - \frac{3}{2}(m_{s}+m_{u})^{2}\left(
         \frac{{\sin}^{2}(\alpha)}{M^{2}_{K_{1}}(1270)}+
         \frac{{\cos}^{2}(\alpha)}{M^{2}_{K_{1}}(1400)}
         \right)\right)^{-1}.
         \end{displaymath}
         
	Here $Z_{\eta_{u}}, Z_{\eta_{s}}$, and $Z_{K}$ are additional renormalization constants appearing in the transitions between axial-vector and pseudoscalar mesons \cite{Volkov:1998ax}, $M_{f_{1}(1285)} = 1282 \pm 0.5$~MeV, $M_{K_{1}(1270)} = 1272 \pm 7$~MeV, $M_{K_{1}(1400)} = 1403 \pm 7$~MeV, $M_{f_{1}(1420)} = 1426.3 \pm 0.9$ MeV \cite{Tanabashi:2018oca}. In the indicated expression for $Z_{K}$, the splitting of the state $K_{1A}$ into two physical mesons $K_{1}(1270)$ and $K_{1}(1400)$ is taken into account \cite{Volkov:2019cja}.
	
	Integrals appearing in quark loops:
	\begin{eqnarray}
		I_{n_{1}n_{2}} =
		-i\frac{N_{c}}{(2\pi)^{4}}\int\frac{\Theta(\Lambda^{2} - k^2)}{(m_{u}^{2} - k^2)^{n_{1}}(m_{s}^{2} - k^2)^{n_{2}}}
		\mathrm{d}^{4}k,
	\end{eqnarray}
	where $\Lambda = 1.25$~GeV is the cutoff parameter \cite{Volkov:1986zb}.
	
	By the ground state of the field $K_{1}$ given in the Lagrangian we mean the field $K_{1A}$ related to nonet $^{3}P_{1}$ and splitting into two physical states $K_{1}(1270)$ and $K_{1}(1400)$ (\ref{K1AK1B}). The contribution of strange axial-vector mesons from the nonet $^{1}P_{1}$  in weak decays can be ignored \cite{Suzuki:1993yc, Volkov:2019cja, Volkov:2019cja}.

\section{The process $\tau \to K^{*-}(892) \eta \nu_{\tau}$ in the NJL model}

\begin{figure}[h]
		\center{\includegraphics[scale = 0.5]{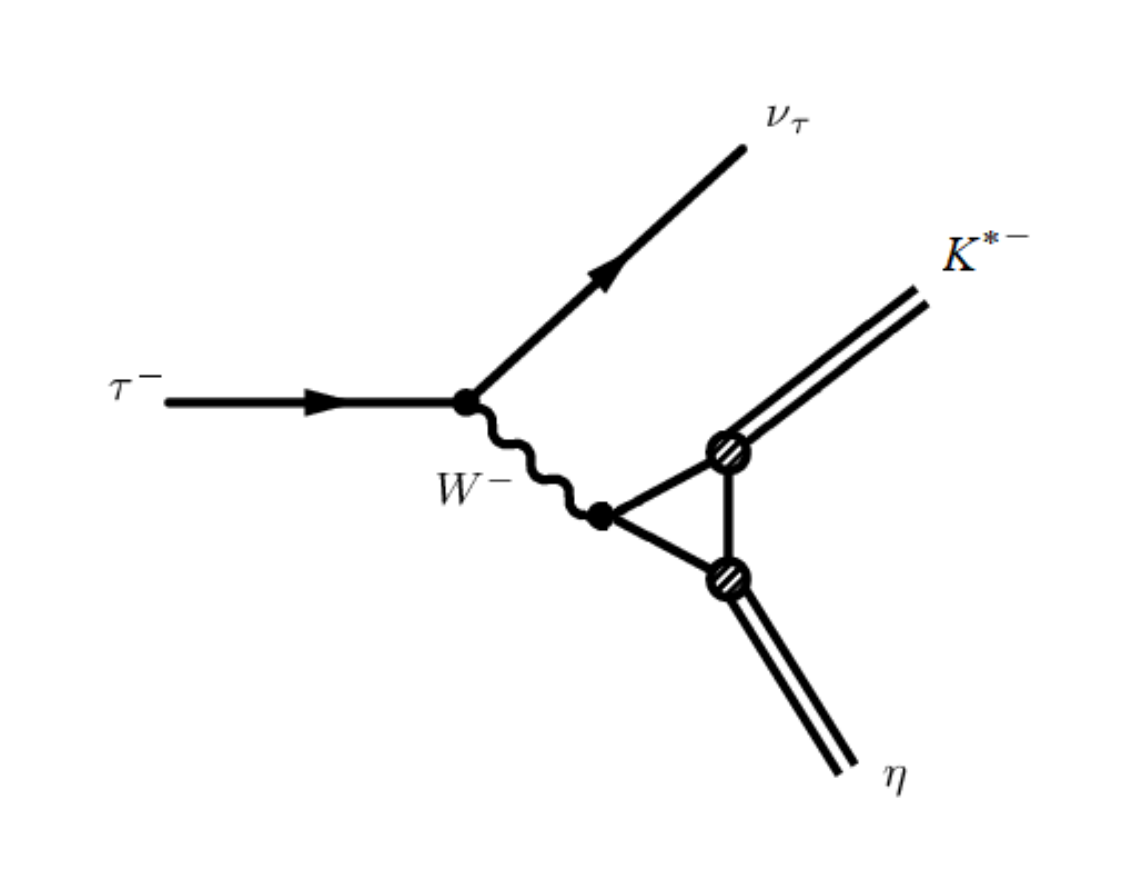}}
		\caption{Contact diagram.}
		\label{Contact}
	\end{figure}

\begin{figure}[h]
		\center{\includegraphics[scale = 0.6]{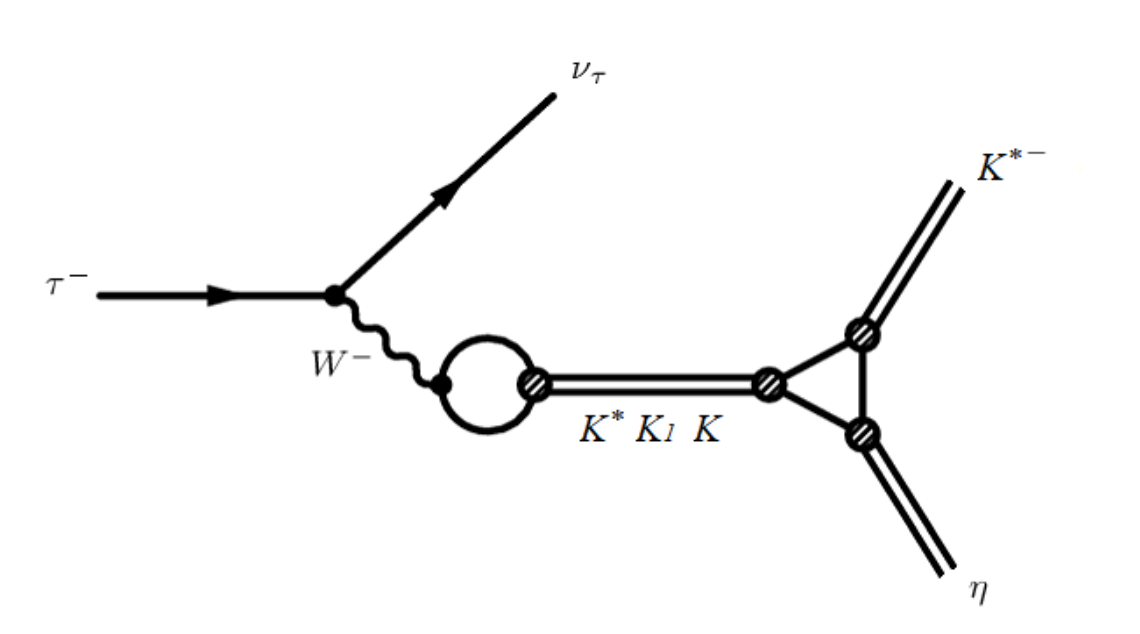}}
		\caption{Diagram with intermediate mesons.}
		\label{Intermediate}
	\end{figure}

The diagrams of the process $\tau \to K^{*-}(892) \eta \nu_{\tau}$ are shown in Figs. ~\ref{Contact} and ~\ref{Intermediate}.
	
	
	The amplitude of the process in the NJL model takes the form:
	\begin{eqnarray}
	&& \mathcal{M}  =  2i G_{F} V_{us} L_{\mu} \biggl[ \mathcal{M}_{c} + \mathcal{M}_{AV(1270)}  \nn \\ && \qquad
	+ \mathcal{M}_{AV(1400)} + \mathcal{M}_{V} + \mathcal{M}_{PS} \biggl]_{\mu\nu} e_{\nu}^{*}(p_{K^{*}}) ,
	\end{eqnarray}
	where $G_{F}$ is the Fermi constant, $V_{us}$ is the element of the Cabibbo-Kobayashi-Maskawa matrix, $L_{\mu}$ is the lepton current and $e_{\nu}^{*}(p_{K^{*}})$ is the polarization vector of the meson $K^{*}(892)$. The terms in parentheses describe the contributions from the contact diagram and from diagrams with various intermediate mesons in the ground states:
	
	\begin{eqnarray}
	&& \mathcal{M}_{c}^{\mu\nu}  =  \frac{3}{2 g_{K^{*}}}\left( m_{s} g_{\eta_{u}} + \sqrt{2} m_{u}g_{\eta_{s}} \right) g^{\mu\nu} \nn \\ && \qquad
	- i \biggl[ m_{u} g_{K^{*}} g_{\eta_{u}} \left[I_{21} + m_{u} (m_{s} - m_{u}) I_{31}\right] \nn \\ && \qquad
         - \sqrt{2} m_{s} g_{K^{*}} g_{\eta_{s}} \left[I_{12} - m_{s} (m_{s} - m_{u}) I_{13}\right] \biggl] \nn \\ && \qquad
         \times e^{\mu\nu\lambda\delta} p_{\eta\lambda} p_{K^{*}\delta},
	\end{eqnarray}	
	\begin{eqnarray}
	&& \mathcal{M}_{AV(1270)}^{\mu\nu}  =  \frac{3}{2 g_{K^{*}}}  \left(m_{s}g_{\eta_{u}} + \sqrt{2} m_{u}g_{\eta_{s}}\right) \nn \\ && \qquad
	\times \biggl[ g^{\mu\lambda}\left[q^{2} - \frac{3}{2}(m_{s} + m_{u})^{2}\right] - q^{\mu}q^{\lambda}\biggl] \nn \\ && \qquad
	\times BW_{\lambda\delta}^{K_{1}(1270)} \sin^{2}(\alpha) g^{\delta\nu},	
	\end{eqnarray}
	\begin{eqnarray}
	&& \mathcal{M}_{AV(1400)}^{\mu\nu}  =  \frac{3}{2 g_{K^{*}}}  \left(m_{s}g_{\eta_{u}} + \sqrt{2} m_{u}g_{\eta_{s}}\right) \nn \\ && \qquad
	\times \biggl[ g^{\mu\lambda}\left[q^{2} - \frac{3}{2}(m_{s} + m_{u})^{2}\right] - q^{\mu}q^{\lambda}\biggl] \nn \\ && \qquad
	\times BW_{\lambda\delta}^{K_{1}(1400)} \cos^{2}(\alpha) g^{\delta\nu}, \nonumber\\
	\end{eqnarray}
	\begin{eqnarray}
	&& \mathcal{M}_{V}^{\mu\nu}  =  -i g_{K^{*}} \biggl[ m_{u} g_{\eta_{u}} \left[I_{21} + m_{u} (m_{s} - m_{u}) I_{31}\right]  \nn \\ && \qquad
	- \sqrt{2} m_{s} g_{\eta_{s}} \left[I_{12} - m_{s} (m_{s} - m_{u}) I_{13}\right]\biggl] \nn \\ && \qquad
	\times \biggl[ g^{\mu\xi}\left[ q^{2} - \frac{3}{2}(m_{s} - m_{u})^{2}\right] - q^{\mu}q^{\xi} \biggl] \nn \\ && \qquad
	\times BW_{\xi\zeta}^{K^{*}} e^{\zeta\nu\lambda\delta} p_{\eta\lambda} p_{K^{*}\delta},
	\end{eqnarray}
	\begin{eqnarray}
	&& \mathcal{M}_{PS}^{\mu\nu}  =  -\frac{3}{2 g_{K^{*}}}(m_{s} + m_{u}) Z_{K} \left(g_{\eta_{u}} + \sqrt{2}g_{\eta_{s}}\right) \nn \\ && \qquad
               \times \biggl[ 1 - \frac{3}{2} (m_{s} + m_{u})^{2} \left(\frac{\sin^{2}(\alpha)}{M_{K_{1}(1270)}^{2}} + \frac{\cos^{2}(\alpha)}{M_{K_{1}(1400)}^{2}}\right) \nn \\ && \qquad
                - \frac{3}{2}\frac{m_{s} g_{\eta^{u}} + \sqrt{2} m_{u} g_{\eta^{s}}}{g_{\eta^{u}} + \sqrt{2} g_{\eta^{s}}}  (m_{s} + m_{u}) \nn \\ && \qquad
                \times \left(\frac{\sin^{2}(\alpha)}{M_{K_{1}(1270)}^{2}} + \frac{\cos^{2}(\alpha)}{M_{K_{1}(1400)}^{2}}\right) \biggl] q^{\mu}q^{\nu}BW^{K}.
	\end{eqnarray}
	
	Intermediate mesons are described by the Breit-Wigner propagators:
	\begin{eqnarray}
	&& BW_{M}^{\mu\nu}  =  \frac{g^{\mu\nu} - \frac{q^{\mu}q^{\nu}}{M_{M}^{2}}}{M_{M}^{2} - q^{2} - i \sqrt{q^{2}} \Gamma_{M}} \nn \\ && \qquad \quad \textrm{(vector and axial-vector case)} 
	\end{eqnarray}
	\begin{eqnarray}
	&& BW_{M}  =  \frac{-1}{M_{M}^{2} - q^{2} - i \sqrt{q^{2}} \Gamma_{M}} \nn \\ && \qquad
	\quad \textrm{(pseudoscalar case)}
	\end{eqnarray}
	
\begin{figure}[h]
\center{\includegraphics[scale = 1.0]{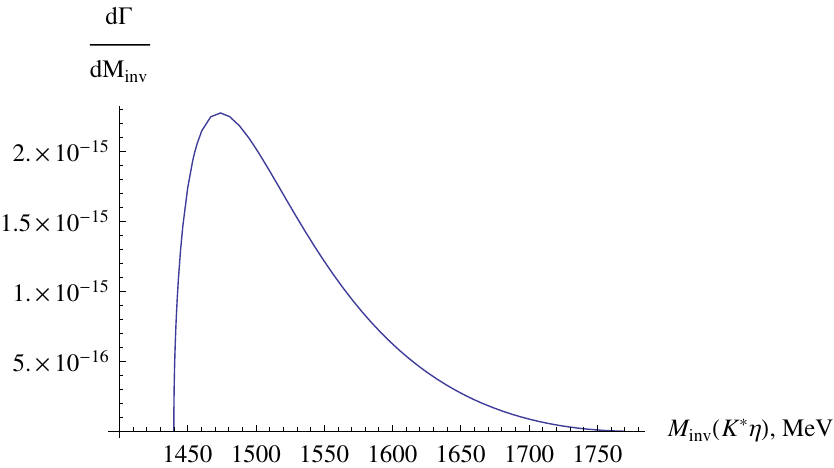}}
\caption{The invariant mass distribution for the system of $K^{*}(892) \eta$ for the decay $\tau \to K^{*}(892) \eta \nu_{\tau}$}
\label{Width}
\end{figure}
	
The results obtained by using this amplitude are shown in Table I.
	
\begin{table}[h!]
\label{Tab}
\begin{center}
\begin{ruledtabular}
\begin{tabular}{cc}
 Channels & Br $(\times 10^{-4})$  \\
 AV & $ 1.217 $  \\
 PS & $ 0.0396 $  \\
 V & $ 0.0028 $  \\
 Total & $ 1.224 $  \\
 \end{tabular}
 \end{ruledtabular}
\end{center}
\caption{Separate contributions to the process $\tau \to K^{*-}(892) \eta \nu_{\tau}$ in the NJL model.}
\end{table}
	
The experimental value for the decay width of this process \cite{Tanabashi:2018oca}:
\begin{eqnarray}
Br(\tau \to K^{*-}(892) \eta \nu_{\tau})_{exp} & = & (1.38 \pm 0.15) \times 10^{-4}.
\end{eqnarray}

\section{Conclusion}
In the present work, a satisfactory description of the decay $\tau \to K^{*}(892) \eta \nu_{\tau}$ was obtained in the framework of the standard NJL model. The calculations show that the main contribution to the decay width $\tau \to K^{*}(892) \eta \nu_{\tau}$ comes from the axial-vector channel with the intermediate mesons $K_{1}(1270)$ and $K_{1}(1400)$ considered together with the contact diagram contribution with the intermediate axial-vector $W$ boson. The contribution from the vector channel with the intermediate $K^{*-}(892)$ meson together with the contact channels with the intermediate vector $W$ boson is 3 orders less than the above contributions. The vector amplitude is orthogonal and does not interfere with other channels. A more significant contribution is made by a channel with an intermediate pseudoscalar $K$ meson. 

As was noted in the Introduction, this process was studied in Ref. \cite{Li:1996md} within the framework of a model close to the NJL one. However, it differs from the NJL model by a slightly different approach to obtain meson vertices, and in particular, vertices with the $W$ boson \cite{Li:1996md, Li:1995tv}. Indeed, for the construction of vertices of mesons with W boson the vector dominance model was used. At the same time, in the NJL model, these vertices are described by using
quark loops like all other meson vertices and the results of the vector dominance model are automatically reproduced.

 Prediction for the differential decay width of $\tau \to K^{*}(892) \eta \nu_{\tau}$ is presented in the Fig.3. It should be emphasized that all results have been obtained with the previously determined parameters of the NJL model without any additional free parameters. The results obtained in this work are in satisfactory agreement with experimental data.

\section*{Acknowledgments}

We are grateful to A.B. Arbuzov for his interest to our work and important remarks which improved the paper. This work is supported by the JINR grant for young scientists and specialists No.19-302-06.


\begin{thebibliography}{32}

    \bibitem{Nambu:1961tp} Y.~Nambu and G.~Jona-Lasinio, Phys.\ Rev.\  {\bf 122}, 345 (1961).
    \bibitem{Eguchi:1976iz} T.~Eguchi, Phys.\ Rev.\ D {\bf 14}, 2755 (1976).
    \bibitem{Ebert:1982pk} D.~Ebert and M.~K.~Volkov, Z.\ Phys.\ C {\bf 16} (1983) 205.
    \bibitem{Volkov:1984kq} M.~K.~Volkov, Annals Phys.\  {\bf 157} (1984) 282.
    \bibitem{Volkov:1986zb} M.~K.~Volkov, Sov. J. Part. Nucl. 17, 186 (1986).
    \bibitem{Ebert:1985kz} D.~Ebert, H.~Reinhardt, Nucl. Phys. B 271, 188 (1986).
    \bibitem{Vogl:1991qt} U.~Vogl and W.~Weise, Prog.\ Part.\ Nucl.\ Phys.\  {\bf 27} (1991) 195.
    \bibitem{Klevansky:1992qe} S.~P.~Klevansky, Rev. Mod. Phys. 64, 649 (1992).
    \bibitem{Volkov:1993jw} M.~K.~Volkov, Phys.\ Part.\ Nucl.\  {\bf 24} (1993) 35.
    \bibitem{Hatsuda:1994pi} T.~Hatsuda and T.~Kunihiro, Phys.\ Rept.\  {\bf 247} (1994) 221.
    \bibitem{Ebert:1994mf} D.~Ebert, H.~Reinhardt, M.~K.~Volkov, Prog. Part. Nucl. Phys 33, 1 (1994).
    \bibitem{Volkov:2005kw} M.~K.~Volkov, A.~E.~Radzhabov, Phys. Usp. 49, 551 (2006).
    \bibitem{Li:1996md} B.~A.~Li, Phys.\ Rev.\ D {\bf 55} (1997) 1436
    \bibitem{Dai:2018thd} L.~R.~Dai, R.~Pavao, S.~Sakai and E.~Oset, Eur.\ Phys.\ J.\ A {\bf 55} (2019) no.2, 20.
    \bibitem{Tanabashi:2018oca} M. Tanabashi et al. (Particle Data Group), Phys. Rev. D {\bf 98}, 030001 (2018).
    \bibitem{Volkov:1998ax} M.~K.~Volkov, M.~Nagy and V.~L.~Yudichev, Nuovo Cim.\ A {\bf 112}, 225 (1999).   
    \bibitem{Volkov:2019fyk} M.~K.~Volkov and A.~A.~Pivovarov, JETP Lett.\  {\bf 110}, no. 4, 237 (2019).   
    \bibitem{Volkov:2019yhy} M.~K.~Volkov, K.~Nurlan and A.~A.~Pivovarov, Int.\ J.\ Mod.\ Phys.\ A {\bf 34} (2019) no.24, 1950137.    
    \bibitem{Li:1995tv} B.~A.~Li, Phys.\ Rev.\ D {\bf 52}, 5184 (1995).
    \bibitem{Volkov:1984fr} M.~K.~Volkov and A.~A.~Osipov, Sov.\ J.\ Nucl.\ Phys.\  {\bf 41} (1985) 500, [Yad.\ Fiz.\  {\bf 41} (1985) 785].
    \bibitem{Suzuki:1993yc} M.~Suzuki, Phys.\ Rev.\ D {\bf 47} (1993) 1252.
    \bibitem{Volkov:2019cja} M.~K.~Volkov, A.~A.~Pivovarov and K.~Nurlan, Eur.\ Phys.\ J.\ A {\bf 55} (2019) no.9,  165.


 \end{thebibliography}
\end{document}